\documentclass[aps,prb,twocolumn,showpacs,superscriptaddress,groupedaddress,10pt]{revtex4-1}

\usepackage{graphicx}
\usepackage{dcolumn}
\usepackage{amsmath}
\usepackage{amssymb}
\usepackage[subrefformat=parens]{subcaption}
\usepackage{physics}

\newcommand{\Nk}{ N_k }

\newcommand{\kk}{\bm{k}}

\newcommand{\cre}[1]{\hat{a}_{#1}^\dagger} 
\newcommand{\ani}[1]{\hat{a}_{#1}}         
\newcommand{\crebar}[1]{\bar{a}_{#1}^\dagger} 
\newcommand{\anibar}[1]{\bar{a}_{#1}}         
\newcommand{\tamp}[2]{t_{#1}^{#2}}         
\newcommand{\lamp}[2]{\lambda_{#1}^{#2}}   
\newcommand{\xamp}[2]{x_{#1}^{#2}}         
\newcommand{\yamp}[2]{y_{#1}^{#2}}         
\newcommand{\Top}{\hat{T}}               
\newcommand{\Lop}{\hat{\Lambda}}           
\newcommand{\Xop}{\hat{X}}                 
\newcommand{\Yop}{\hat{Y}}                 
\newcommand{\hamil}{\mathcal{H}}           
\newcommand{\hamilbar}{\bar{\mathcal{H}}}  
\newcommand{\EHF}{ E_\mathrm{HF}}
\newcommand{\ECCSD}{ E_\mathrm{CCSD}}

\newcommand{\affil}{Department of Applied Physics, The University of Tokyo, Tokyo 113-8656, Japan}

\begin{document}

\title{Band structures in coupled-cluster singles-and-doubles Green's function (GFCCSD)}
\author{Yoritaka Furukawa} \affiliation{\affil}
\author{Taichi Kosugi} \affiliation{\affil}
\author{Hirofumi Nishi} \affiliation{\affil}
\author{Yu-ichiro Matsushita} \affiliation{\affil} 
\date{\today}

\begin{abstract}
We demonstrate that coupled-cluster singles-and-doubles Green's function (GFCCSD) method is a powerful and prominent tool drawing the electronic band structures and the total energies,
which many theoretical techniques struggle to reproduce.
We have calculated single-electron energy spectra
via GFCCSD method for various kinds of systems, ranging from ionic to covalent and van der Waals, for the first time: one-dimensional LiH chain, one-dimensional C chain, and one-dimensional Be chain.
We have found that the band gap becomes narrower than in HF due to the correlation effect. We also show that the band structures obtained from GFCCSD method include both quasiparticle and satellite peaks successfully.
Besides, taking one-dimensional LiH as an example, we discuss the validity of restricting the active space to suppress the computational cost of GFCCSD method while maintaining the accuracy.
We show that the calculated results without bands that do not contribute to the chemical bonds are in good agreement with full-band calculations.
With GFCCSD method, we can calculate the total energy and band structures with high precision.
\end{abstract}

\maketitle

\section{Introduction}
Construction of a novel advanced calculation methodology for high accuracy is always one of the most important themes in theoretical materials science.
One of the most successful theories in this context must be the density-functional theory (DFT) \cite{HK,KS}.
The DFT has been applied to a wide variety of systems, from finite to extended, and enables us to reproduce structural parameters such as lattice constants within a few percent of error and even to predict the material properties with relatively cheap computational cost.
According to Janak's theorem \cite{janak1978} with Kohn-Sham (KS) equation in the DFT framework, we can draw one-electron energy levels for finite systems and electronic band structures for periodic systems. This is another noteworthy property of the DFT because electric energy levels and band structures can be experimentally observed through X-ray photoelectron spectroscopy (XPS) and angle-resolved photo-emission spectroscopy (ARPES) \cite{damascelli2003}.
Comparison of the band structures obtained from ARPES measurements and DFT calculations helps us deepen our understandings of the electronic properties of materials.

Behind the great successes of the DFT, we start to notice some drawbacks in the DFT at the same time. Well-known examples are that the DFT cannot reproduce van der Waals interaction, satellite peaks, and Mott gaps.
Many efforts have been done so far to solve such difficulties, including
self-interaction correction (SIC) method \cite{SICDFT},
LDA+U \cite{LDAU},
hybrid functionals \cite{hybrid1,hybrid2,hybrid3},
LDA+DMFT \cite{DMFT1,DMFT2,DMFT3},
GW \cite{GW1,GW2,GW3},
GW+cumulant expansion \cite{GWC1,GWC2},
van der Waals DFT \cite{DFTvdW1,DFTvdW2},
RDMFT \cite{RDMFT1,RDMFT2,Sangeeta}, etc.

From the viewpoint of the development of a methodology, the wave function theory (WFT) has a great advantage in comparison with the DFT. One can improve the accuracy relatively easily within the WFT, while it is difficult in the DFT.
However, the application of the WFT has been mostly limited to finite systems so far due to the huge calculation cost.
Very recently, owing to the dramatical development of supercomputers, some groups have succeessfully demonstrated the application of the WFTs to periodic systems.
For example, density-matrix-renormalization group (DMRG) \cite{DMRG1,DMRG2}, the transcorrelated method \cite{TC1,TC2,TC3,TC4,TC5,TC6}, and the Monte-Carlo configuration interaction　\cite{FCIQMC,NiOCCSD} have been reported to be applied to periodic systems.
Most previous studies with the WFT, however, focused only on the ground state energies except for the transcorrelated method. For the most standard WFTs, drawing electronic band structures is not trivial.

Among WFTs, coupled-cluster theory \cite{Monkhorst1977,Stanton1993,Bartlett2007} is known to be a highly successful scheme that is capable of efficiently incorporating electronic correlations.
Coupled-cluster singles and doubles (CCSD) method, which expands the reference state using single and double excitation operators, is the most popular type of the implementations due to its high accuracy and computational feasibility.
CCSD method has been applied to periodic systems of strongly-correlated systems such as NiO \cite{NiOCCSD}.
However, CCSD method cannot draw the single-electron energy spectrum in the standard form, along with other WFTs.
Electronic excited states can be also calculated in CC theory by using the equation-of-motion CC (EOM-CC) \cite{Monkhorst1977,Stanton1993} or the symmetry-adapted cluster/configuration interaction (SAC-CI) \cite{SACCIPk} method.
EOM-CCSD has already used for silicon crystal \cite{mcclain2017}.
Recently, a method to obtain one-body Green's functions based on CC theory (GFCC) was proposed \cite{Nooijen92,Nooijen93,Nooijen95}, with which one can obtain the one-electron energy spectrum of materials.
It has been, however, only applied to a limited number of systems. In particular, no periodic system has ever been treated by GFCCSD method except for homogeneous electron gas \cite{mcclain2017}.

In this work, we have calculated band structures of several kinds of materials, ranging from ionic to covalent and van der Waals systems, through GFCCSD method.
We have found that GFCCSD method is a powerful and prominent tool drawing the electronic band structures and yielding total energy at one time by demonstrating the results.
We present the calculation results of periodic systems, which are one-dimensional LiH chain, C chain, and Be chain.
We also show the band structures obtained from GFCCSD calculations for the first time, in which we see the emergence of satellite peaks. We also discuss how the calculations are affected by the reduction of active space, which is an important factor in reducing the computational cost.

\section{Green's function from the coupled-cluster calculations} \label{sec:method}

The present study is restricted only to the non-relativistic Hamiltonian, $\hamil$. In the coupled-cluster theory, the ground state wave function $\ket{\Psi_\mathrm{CC}}$ is described to be
\begin{equation}
    \ket{\Psi_\mathrm{CC}} = e^{\Top} \ket{\Psi_0},
\label{wfn_CCSD}
\end{equation}
where $\ket{\Psi_0}$ is a so-called reference state, which usually adopts the Hartree--Fock wave function. The operator $\Top$ represents the $p$-electron excitation and is defined as
\begin{eqnarray}
    \Top_p =
    \frac{1}{(p!)^2}
    \sum_{
    \substack{i,j,k,\dots, \kk_i\kk_j\kk_k\dots \\
              a,b,c,\dots, \kk_a\kk_b\kk_c\dots
             }
    }
    \tamp{i\kk_i j\kk_j k\kk_k \dots}{a\kk_a b\kk_b c\kk_c \dots} \cdot \\
    \cre{a\kk_a} \cre{b\kk_b} \cre{c\kk_c}
    \cdots
    \ani{k\kk_k} \ani{j\kk_j} \ani{i\kk_i}
\end{eqnarray}

where $\cre{p \kk_p}$ and $\ani{p \kk_p}$ are creation and annihilation operators of an electron with momentum $\kk_p$ at state $p$, respectively.
The indices $i,j,\cdots$ ($a,b,\cdots$) represent occupied (unnoccupied) states, whereas $p,q,\cdots$ are used for any states, regardless of whether they are occupied or unnoccupied ones.
The coefficients in $\Top$, $\tamp{i\kk_i j\kk_j k\kk_k \dots}{a\kk_a b\kk_b c\kk_c \dots}$, are determined from the amplitude equations, which are deduced by projecting excited states
$\bra{\Psi_{i\kk_ij\kk_j\cdots}^{a\kk_ab\kk_b\cdots}}$
to the Schr\"odinger equation $\hamil \ket{\Psi} = E\ket{\Psi}$, in which a similarity transformed Hamiltonian $\hamilbar = e^{-\Top} \hamil e^{\Top}$ appears:
\begin{equation}
    \mel{ \Psi_{i\kk_ij\kk_j\cdots}^{a\kk_ab\kk_b\cdots} }{ \hamilbar }{\Psi_0} =0 \ .
\end{equation}
After determining the coefficients in $\Top$, the total energy $\ECCSD$ can be calculated by projecting $\bra{\Psi_0}$:
\begin{equation}
    \ECCSD = \mel{\Psi_0}{e^{-\Top} \hamil e^{\Top}}{\Psi_0} .
\end{equation}

One-particle Green's function of the frequency representation at zero temperature is written as
\begin{equation}
\begin{split}
    G_{p\kk_pq\kk_q}(\omega)
    = &G_{p\kk_pq\kk_q}^{(h)}(\omega) + G_{p\kk_pq\kk_q}^{(e)}(\omega) \\
    =
    &\mel{\Psi}{ \cre{q\kk_q} \frac{1}{\omega+\hamil_{N}} \ani{p\kk_p} }{\Psi}\\
    &+\mel{\Psi}{ \ani{q\kk_q} \frac{1}{\omega-\hamil_{N}} \cre{p\kk_p} }{\Psi},
\label{def_green}
\end{split}
\end{equation}
in which the Green's function is separated into the electron removal and attachment part (partial Green's functions).
The $\hamil_{N}$ is defined as $\hamil_{N} = \hamil - E_0$, where $E_0$ is the total energy of the exact ground state $\ket{\Psi}$. Here, one adopts the CCSD wave function to the exact wave function, $\ket{\Psi}=\ket{\Psi_\mathrm{CC}}$. Using the similarity transformed Hamiltonian
$\hamilbar_{N} = e^{-\Top} \hamil e^{\Top} - E_0$
and the transformed creation and annihilation operators
$\crebar{q\kk_q} = e^{-\Top} \cre{q\kk_q} e^{\Top}$
and
$\anibar{p\kk_p} = e^{-\Top} \ani{p\kk_p} e^{\Top}$
, we can rewrite the partial Green's functions to
\begin{equation}
    G_{p\kk_pq\kk_q}^{(h)}(\omega) =
    \mel{\Psi_0}{ (1+\Lop) \crebar{p\kk_p} \frac{1}{\omega+\hamilbar_N} \anibar{q\kk_q} }{\Psi_0},
\end{equation}
\begin{equation}
    G_{p\kk_pq\kk_q}^{(e)}(\omega) =
    \mel{\Psi_0}{ (1+\Lop) \anibar{p\kk_p} \frac{1}{\omega-\hamilbar_N} \crebar{q\kk_q}}{\Psi_0}.
\end{equation}
Note that the transformed Hamiltonian $\hamilbar_{N}$ is not Hermitian and that the Green's function is constructed using bi-variational method \cite{Arponen83,Stanton1993,Bi-vari2}.
The operator $\Lop$ is a de-excitation operator which is determined by solving
\begin{equation}
    \mel{ \Psi_{i\kk_ij\kk_j\cdots}^{a\kk_ab\kk_b\cdots} }
        { (1+\Lop) e^{-\Top} \hamil e^{\Top} }
        {\Psi_0} = 0.
\end{equation}

In order to avoid the computational difficulty in treating the inverse matrix
$(\omega \pm \hamilbar_N)^{-1}$,
$\Xop_{q\kk_q}(\omega)$ and $\Yop_{q\kk_q}(\omega)$ are introduced as follows:
\begin{equation}
    (\omega+\hamilbar_N) \Xop_{q\kk_q} (\omega) \ket{\Psi_0} =
    \ani{q\kk_q} \ket{\Psi_0},
    \label{eq:ipgf}
\end{equation}
\begin{equation}
    (\omega-\hamilbar_N)\Yop_{q\kk_q}(\omega)\ket{\Psi_0} =
    \cre{q\kk_q} \ket{\Psi_0}.
    \label{eq:eagf}
\end{equation}
Once we solve Eq.~(\ref{eq:ipgf}) and (\ref{eq:eagf}), we can get the information of (N$-1$)- and (N$+1$)-electron states involved in the Green's function, respectively.
Note that these two linear equations are equivalent to Hamiltonian of EOM-CC theory: Eq.~(\ref{eq:ipgf}) corresponds to (N$-1$)-electron states yielding ionization potential (IP-EOM-CC) and Eq.~(\ref{eq:eagf}) corresponds to (N$+1$)-electron states (EA-EOM-CC).
With $\Xop_{q\kk_q} (\omega)$ and $\Yop_{q\kk_q} (\omega)$, the Green's function is finally expressed as \cite{Kowalski2014,Kowalski2016}
\begin{equation}
    G_{p\kk_pq\kk_q}^{(h)}(\omega) =
    \mel{\Psi_0}{ (1+\Lop) \crebar{p\kk_p} \Xop_{q\kk_q} (\omega) }{\Psi_0},
\end{equation}
\begin{equation}
    G_{p\kk_pq\kk_q}^{(e)}(\omega) =
    \mel{\Psi_0}{ (1+\Lop) \anibar{p\kk_p} \Yop_{q\kk_q} (\omega) }{\Psi_0}.
\end{equation}
We can calculate single-electron spectra $A(\omega)$ using the Green's function:
\begin{equation}
    A(\omega) =
    - \frac{1}{\pi} \Im
    \left[
        \tr \left( G(\omega+i\delta) \right)
    \right].
\label{spectr}
\end{equation}

The band structure is obtained simply by decomposing $A(\omega)$ into the contributions from each $k$-point, $A_{\bm{k}}(\omega)$:
\begin{equation}
A_{\bm{k}}(\omega) =
    - \frac{1}{\pi} \Im
    \left[
        \sum_p G_{p\kk_pp\kk_p}(\omega+i\delta)
    \right]
\end{equation}

In this study, we truncate the excitation operator $\Top$ up to singles and doubles (CCSD) as follows:
\begin{eqnarray}
    \Top &\simeq&
    \sum_{i\kk_ia\kk_a} \tamp{i\kk_i}{a\kk_a} \cre{a\kk_a} \ani{i\kk_i} \nonumber \\
    &+&
    \frac{1}{4} \sum_{i\kk_ij\kk_ja\kk_ab\kk_b} \tamp{i\kk_ij\kk_j}{a\kk_ab\kk_b} \cre{a\kk_a} \cre{b\kk_b} \ani{j\kk_j} \ani{i\kk_i}.
\end{eqnarray}
By introducing the truncation in the $\Top$ operator, we derive the following equations for $\Lop$, $\Xop_{q\kk_q}$, and $\Yop_{q\kk_q}$ operators maintaining the same accuracy as CCSD:
\begin{eqnarray}
    \Lop &\simeq&
    \sum_{i\kk_ia\kk_a} \lamp{i\kk_i}{a\kk_a} \cre{i\kk_i} \ani{a\kk_a} \nonumber \\
    &+&
    \frac{1}{4} \sum_{i\kk_ij\kk_ja\kk_ab\kk_b} \lamp{i\kk_ij\kk_j}{a\kk_ab\kk_b} \cre{i\kk_i} \cre{j\kk_j} \ani{b\kk_b} \ani{a\kk_a}
\end{eqnarray}
\begin{eqnarray}
    \Xop_{q\kk_q}(\omega) &\simeq&
    \sum_{i\kk_i} \xamp{i(q\kk_q)}{ } (\omega) \ani{i\kk_i} \nonumber \\
    &+&
    \frac{1}{2} \sum_{i\kk_ij\kk_ja\kk_a} \xamp{i\kk_ij\kk_j(q\kk_q)}{a\kk_a} (\omega) \cre{a\kk_a} \ani{j\kk_j} \ani{i\kk_i}
\end{eqnarray}
\begin{eqnarray}
    \Yop_{q\kk_q}(\omega) &\simeq&
    \sum_{a\kk_a}y_{a\kk_a(q\kk_q)}(\omega)\cre{a\kk_a} \nonumber \\
    &+&
    \frac{1}{2} \sum_{i\kk_ia\kk_ab\kk_b} \yamp{i\kk_i(q\kk_q)}{a\kk_ab\kk_b} (\omega) \cre{a\kk_a} \cre{b\kk_b} \ani{i\kk_i}.
\end{eqnarray}
In particular, $\Xop_{q\kk_q}$ operators are truncated up to $1h$ (first term of the right hand side (r.h.s.)) and $2h1p$ term (second term of the r.h.s.), and $\Yop_{q\kk_q}$ operators are similarly truncated up to $1p$ (first term of the r.h.s.) and $2p1h$ term (second term of the r.h.s.). These truncation for $\Xop_{q\kk_q}$ and $\Yop_{q\kk_q}$ leads to the expression of the wave function after electron attachment/removal to be
\begin{equation}
\begin{split}
    \ket{\Psi^{N-1}_{q\kk_q}}
    =
    &e^{\Top} \sum_{i\kk_i} \xamp{{i\kk_i}(q\kk_q)}{} (\omega) \ani{i\kk_i} \ket{\Psi_0} \\
    &+
    e^{\Top} \sum_{i\kk_ij\kk_ja\kk_a} \xamp{i\kk_ij\kk_j(q\kk_q)}{a\kk_a} (\omega) \cre{a\kk_a} \ani{j\kk_j} \ani{i\kk_i} \ket{\Psi_0} \\
    \equiv
    &e^{\Top} \sum_{1h} \ket{1h} + e^{\Top} \sum_{2h1p} \ket{2h1p}
    \label{def_Psi_N-1}
\end{split}
\end{equation}

\begin{equation}
\begin{split}
    \ket{\Psi^{N+1}_{q\kk_q}}
    =
    &e^{\Top} \sum_{a\kk_a} \yamp{a\kk_a(q\kk_q)}{} (\omega) \cre{a\kk_a} \ket{\Psi_0}\\
    &+
    e^{\Top} \sum_{i\kk_ia\kk_ab\kk_b} \yamp{i\kk_i(q\kk_q)}{a\kk_ab\kk_b} (\omega) \cre{a\kk_a} \cre{b\kk_b} \ani{i\kk_i} \ket{\Psi_0} \\
    \equiv
    &e^{\Top} \sum_{1p} \ket{1p} + e^{\Top} \sum_{2p1h} \ket{2p1h},
    \label{def_Psi_N+1}
\end{split}
\end{equation}
where we introduced notations describing subspace in Hilbert space, $\ket{1h}$, $\ket{2h1p}$, $\ket{1p}$, and $\ket{2p1h}$, representing one electron annihilated, one electron annihilated and one electron excited, one electron created, and one electron created and one electron excited from the HF electron configuration, respectively.

The computational cost for CCSD and $\Lambda$-CCSD is $\order{N^6\Nk^4}$, where $\Nk$ is the number of sampled $k$-points in the Brillouin zone.
Solving the IP/EA-EOM-CCSD linear equations is computationally demanding.
We use the LU-decomposition method, which costs $\order{N^9\Nk^6 N_{\omega}}$, where $N_{\omega}$ is the number of $\omega$ mesh.

\section{Results} \label{sec:periodic_1d}
\subsection{One-dimensional LiH chain}\label{results_LiH}

We first show the calculated results of one-dimensional LiH chain.
We consider a system in which Li and H atoms are aligned alternately and the Li-H bond lengths are the same everywhere.

\begin{figure}[tb]
   \begin{center}
       \includegraphics[width=0.7\linewidth]{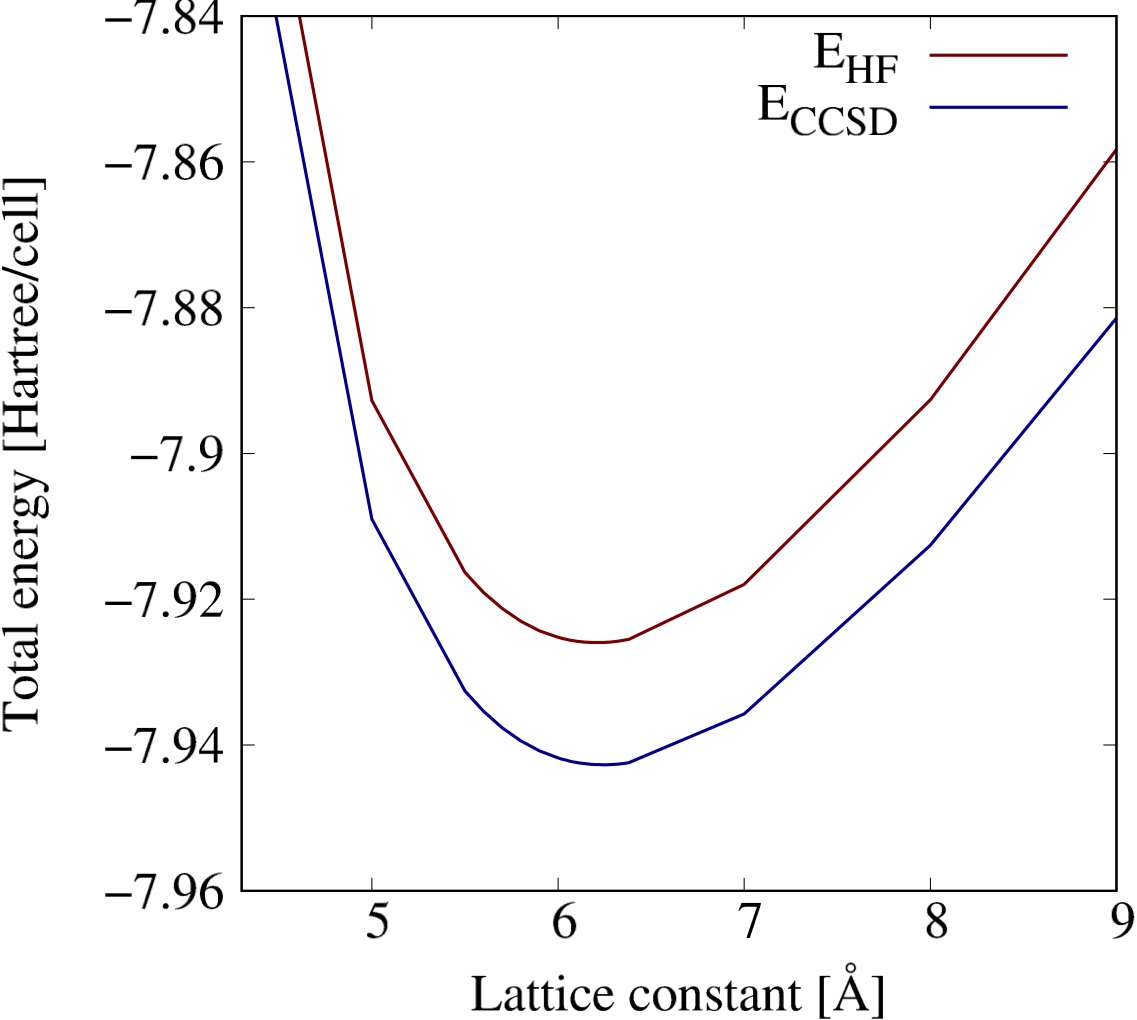}
       \caption{Dependence of the total energy of LiH chain on the lattice constant. Red and blue line represent the total energy obtained from HF and CCSD calculations, respectively.
       \label{img:LiH-Etot-lattice}}
   \end{center}
\end{figure}

We first optimized the lattice constant based on HF and CCSD.
The reference state in Eq.~(\ref{wfn_CCSD}) has been obtained by the restricted Hartree--Fock (RHF) method with the STO-3G basis set, i.e., H-$1s$, Li-$1s$, Li-$2s$, and Li-$2p$ orbitals. The number of sampling $k$-points, whom we shall refer to as $\Nk$ throughout this paper, is set to be 8 for this examination.
In Fig.~\ref{img:LiH-Etot-lattice}, we show the total energy from HF calculations, $\EHF$, and that from CCSD calculations, $\ECCSD$.
We find that the total energy is minimized at $6.21$ {\AA} in HF and $6.24$ {\AA} in CCSD calculations. Here we note that both the lattice constant and the minimized total energy in HF scheme are comparable to those calculated in the past studies \cite{shukla1998, delhalle1980}. We hereby adopt the latter one to be the lattice constant of LiH chain throughout this paper.

Next we determined $\Nk$ by checking the dependence of the lattice constant and the band structures on $\Nk$. We have compared the optimized lattice constant with $\Nk=8$ and that with $\Nk=16$.
We have found that the two lattice constants do not change within $0.01$ {\AA} difference. This shows that $\Nk=8$ is large enough for the calculations of the lattice constant.
Therefore, we adopted $\Nk=8$ in the subsequent calculations.

\begin{figure}
    \includegraphics[width=0.9\linewidth]{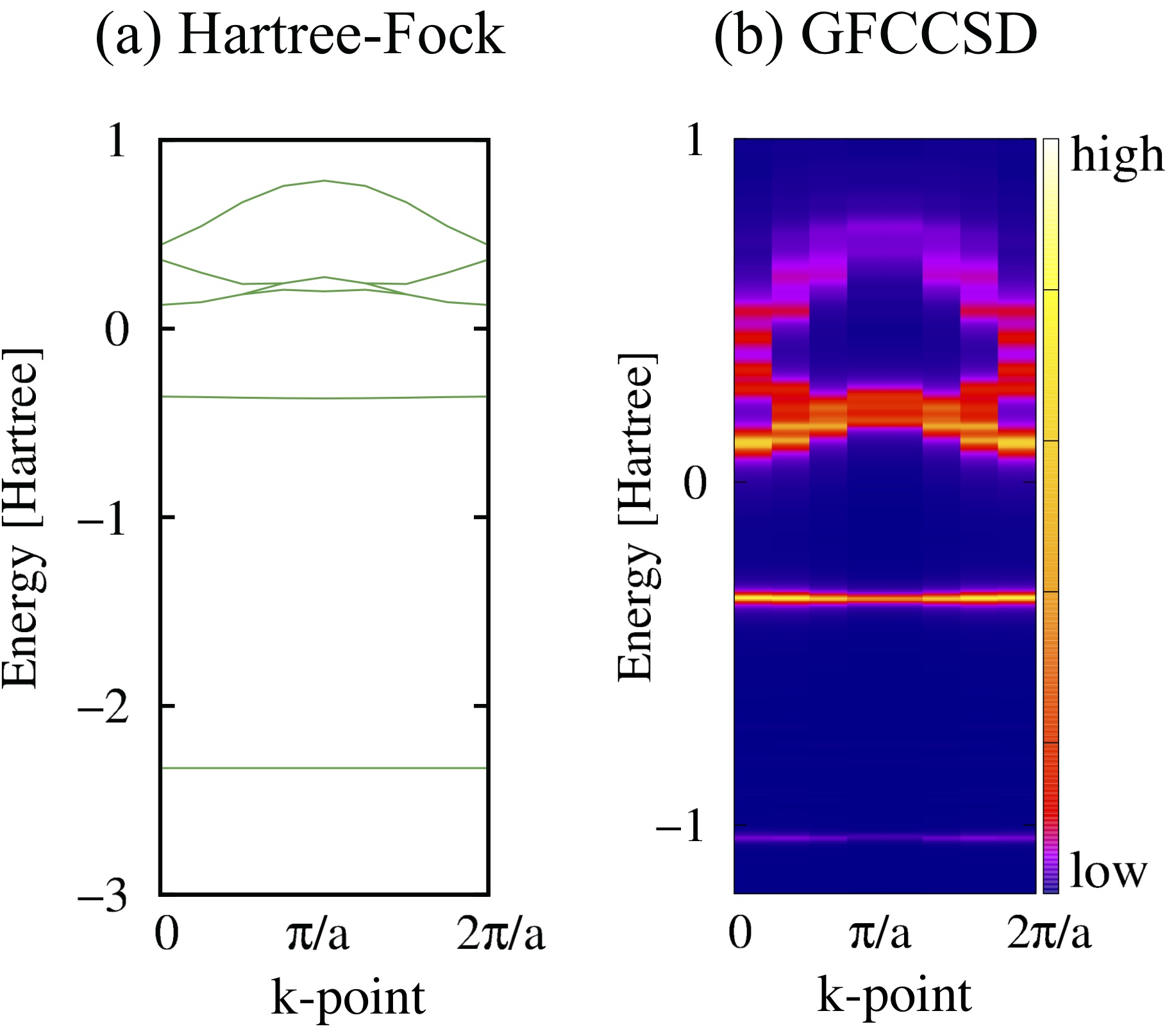}
    \caption{Band structure of the LiH chain from HF (a) and GFCCSD (b) calculated with $\Nk=8$. $a$ in the horizontal axis represents the lattice constant.}
    \label{LiH-spectrum-all}
\end{figure}

Fig.~\ref{LiH-spectrum-all}(a) shows the calculated band structure by HF method.
Two valence and four conduction bands appear, all of which are spin-degenerate.
The system is a typical ionic one, and an electron is thought to be transferred from Li to H atom. By checking the wave function character of each band, we have confirmed that the lowest band at $-2.3$ Hartree is mainly attributed to Li-$1s$ orbital while the second lowest one to H-$1s$ orbital.
The lower two conduction bands are made up of Li-$2p$ orbitals whose directions are orthogonal to the Li-H bonds.
The third lowest band is Li-$2s$ orbital, while the highest energy band is Li-$2p$ orbital. The calculated band gap at the $\Gamma$ point is $0.49$ Hartree.
We present the band structure in GFCCSD scheme in Fig.~\ref{LiH-spectrum-all}(b) with $\delta=0.005$ Hartree.
Compared with the band structure in HF scheme, its quasiparticle bands have become broad especially in the conduction bands representing the finite lifetime of quasiparticles.
The calculated band gap is $0.45$ Hartree in GFCCSD, which is narrower than that in HF.
This fact agrees with the empirical rule that the correlation effect narrows band gaps. Another striking feature is the emergence of satellite bands at $-1.04$ Hartree.

\begin{figure}
    \begin{minipage}{1\linewidth}
        \centering
        \subcaption{Overall peaks}
        \includegraphics[width=0.7\linewidth]{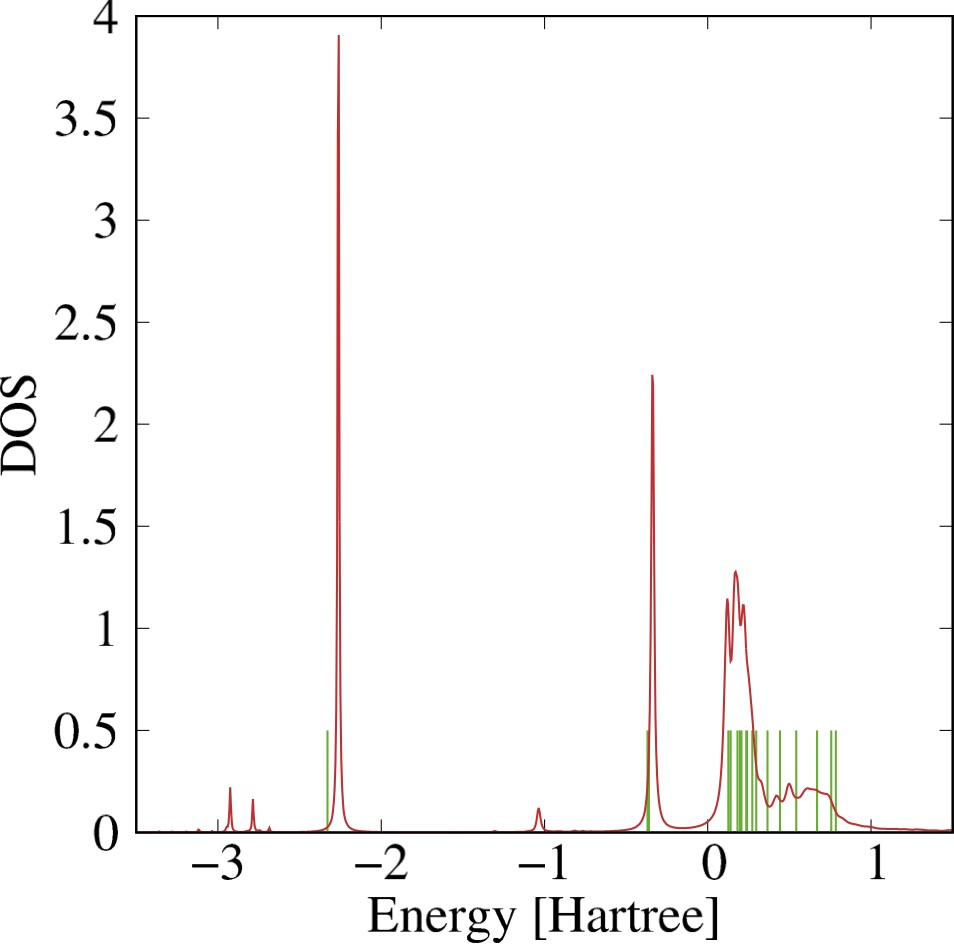}
    \end{minipage}\\
    ~\\~\\~
    \begin{minipage}{1\linewidth}
        \centering
        \subcaption{Satellite peaks}
        \includegraphics[width=0.7\linewidth]{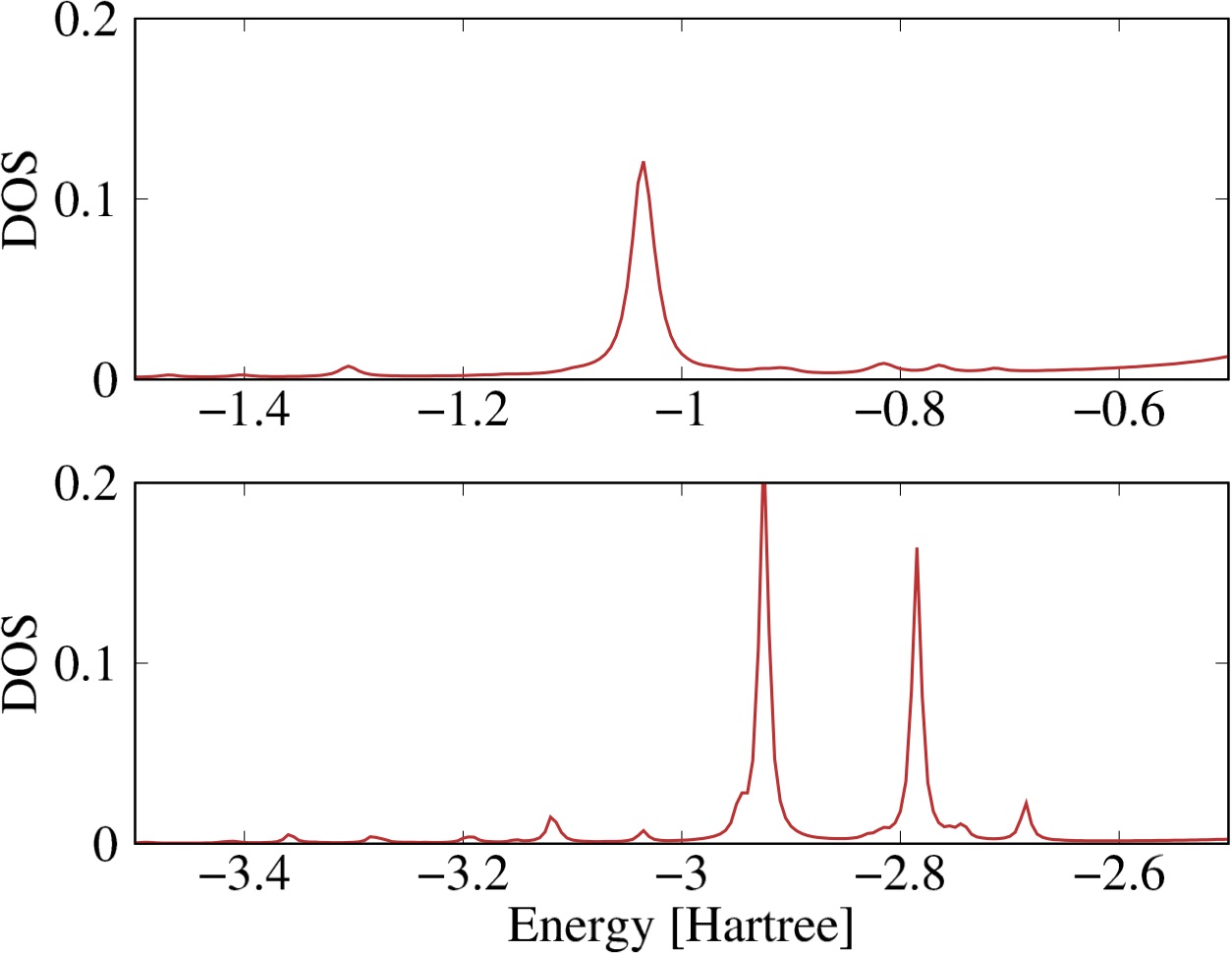}
    \end{minipage}
    \caption{The density of states (DOS) of LiH chain calculated from GFCCSD with $\Nk=8$. The positions of the green sticks in (a) represent the eigenvalues obtained in HF calculations. In (a), its overall shape are shown. The regions where satellite peaks emerge are enlarged in (b).
    \label{img:LiH-DOS-k8}}
\end{figure}

Fig.~\ref{img:LiH-DOS-k8} shows the density of states (DOS) of the LiH chain. It should be noted that the DOS in Fig.~\ref{img:LiH-DOS-k8} shows very spiky peaks due to the limitation of the $k$-point sampling. It is expected, therefore, that at $\Nk \to \infty$ limit the gaps between the spiky peaks become smaller to be a continuous spectrum.
In the plot, we observe two sharp peaks at $-2.30$ and $-0.34$ Hartee, broader peaks near $0.1$ Hartree, and a hump-like one located at around $0.6$ Hartree. We identify that all of these are quasiparticle peaks that correspond to certain energy bands.
We can identify the characters of these peaks by checking the wave function at each peak.
The two sharp peaks correspond to the lowest and the second lowest bands.
The group of peaks derive from the three conduction bands that are located in the range between $0$ and $0.3$ Hartree in Fig.~\ref{LiH-spectrum-all}. The hump-like peak corresponds to the highest conduction band.
We also confirm the shift of these peaks from the HF results, which are indicated by green sticks in Fig.~\ref{img:LiH-DOS-k8}. The lowest and second lowest quasiparticle peaks in GFCCSD are about $0.07$ and $0.03$ Hartree higher than in HF scheme each, while the conduction-band minimum is lower by $0.01$ Hartree.
The energy position of the satellite peaks We observed are at $-2.93$, $-2.79$, and $-1.04$ Hartree. Above 0 Hartree, in contrast, we find no clear satellite peaks.
By integrating the satellite peaks between the first and second peaks, the weight is calculated to be 0.14.

\subsubsection{Restricting the active space in LiH chain} \label{subsec:active-space}

Since the calculation cost of GFCCSD is huge, which is at least $\order{N^6\Nk^4}$ for periodic systems, the number of orbitals to take into account should be suppressed, or minimize the size of the active space in other words, as long as the accuracy of the calculation is maintained.
One idea is to exclude some orbitals that are unlikely to improve the reference wave functions, such as deep levels or unoccupied orbitals that are far from the Fermi levels.
This consideration is what is called the restriction of the active space in quantum chemistry.
This has to be done carefully by considering the physical meaning of each orbital in the material.

\begin{figure}[tb]
    \begin{center}
        \includegraphics[width=0.7\linewidth]{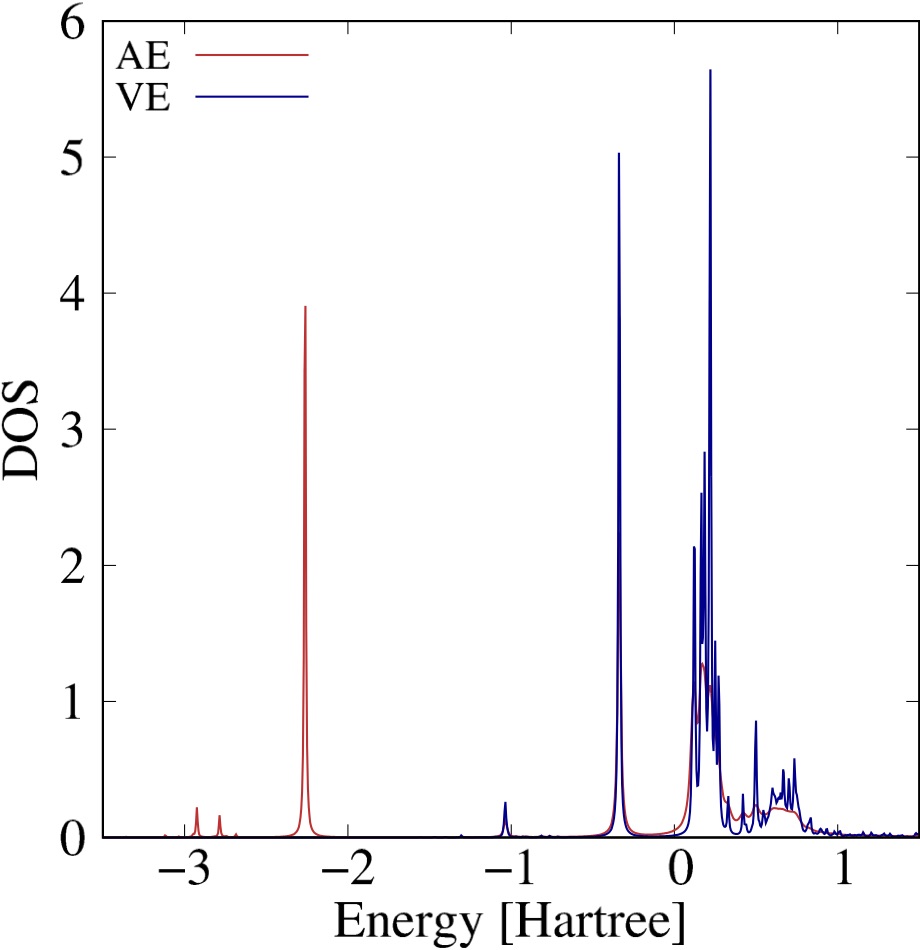}
        \caption{Comparison of all-electron (AE) and valence-electron (VE) calculations of LiH chain for DOS.
        \label{img:LiH-AE-VE-comparison}}
    \end{center}
\end{figure}

To check the validity of the choice of the active space in the LiH chain case, we first examined the DOS with only changing the active space from subsec.\ref{results_LiH}.
As shown in Fig.~\ref{LiH-spectrum-all}(a), of the two valence bands, the lower one is energetically far from the Fermi energy, implying that its contribution to the correlation energy might be negligible.
Therefore, we performed GFCCSD calculations neglecting the lowest band. We compared the DOS from that in subsec.~\ref{results_LiH} and that with the smaller active space, which is presented in Fig.~\ref{img:LiH-AE-VE-comparison}.
We find that the peak positions are identical to each other above $-1.5$ Hartree. The VE plot shown as a blue line has no peak below $-2.0$ Hartree because of the lack of the lowest band in its active space. This manifests that by choosing the proper active space, we can reduce the calculation cost without reducing the calculation accuracy.

\subsection{One-dimensional C chain}

\begin{figure}[tb]
    \begin{minipage}{1\linewidth}
        \centering
        \subcaption{Hartree--Fock}
        \includegraphics[width=0.7\linewidth]{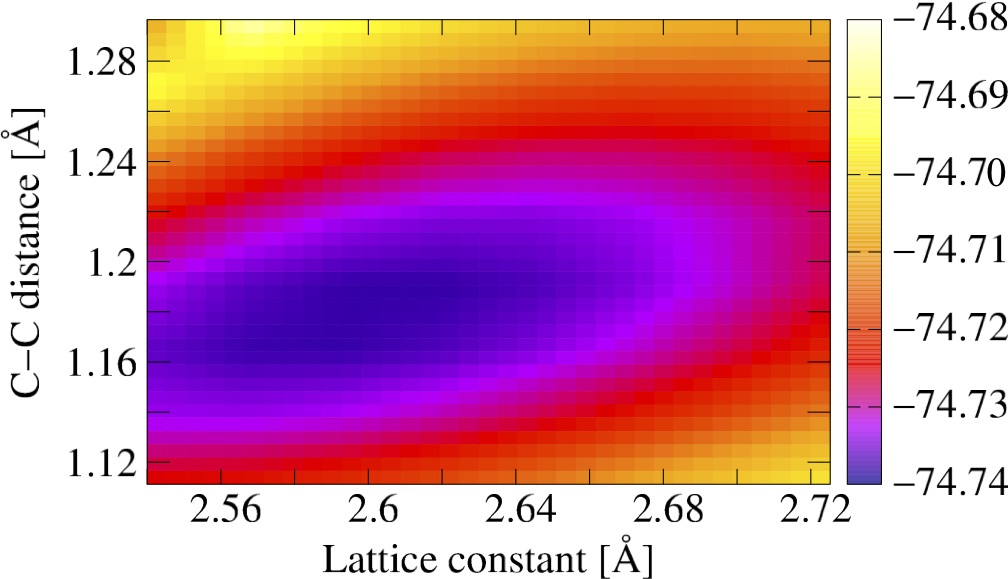}
    \end{minipage}\\
    ~\\~\\~
    \begin{minipage}{1\linewidth}
        \centering
        \subcaption{CCSD}
        \includegraphics[width=0.7\linewidth]{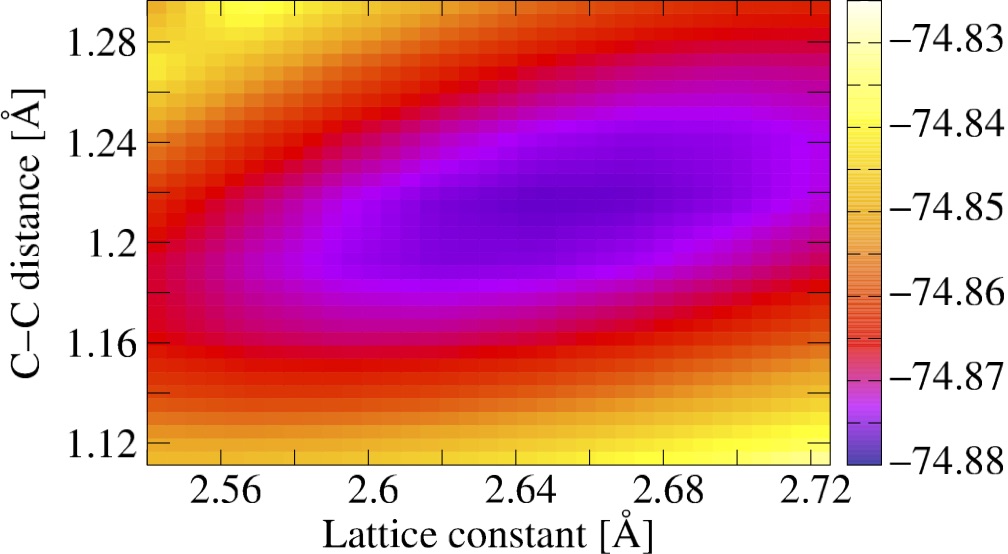}
    \end{minipage}
    \caption{Dependence of the total energy of C chain on the lattice constant (horizontal axis) and the C-C distance within a unit cell (vertical axis) calculated with $\Nk=4$. The energy is shown in the unit of Hartree.
    \label{img:C-Etot}}
\end{figure}

The unit cell of a C chain contains two inequivalent C atoms to form periodically arranged dimers.
The geometric structure of C chain has been determined to be the one that minimizes the CCSD total energy. We have relaxed both the lattice constant and the C-C bond lengths at the same time with the STO-3G basis set, which includes C-$1s$, $2s$, $2p$ orbitals.
The energy surface is shown in Fig.~\ref{img:C-Etot}.
The optimized lattice constant and the C-C bond length have been found to be $5.0$ and $2.29$ Bohr ($2.65$ and $1.21$ {\AA}), respectively.
These values are 5\% larger and 1\% smaller than experimental ones, $4.76$ and $2.32$ Bohr ($2.52$ and $1.23$ {\AA}) \cite{shi2016}.

\begin{figure}
    \includegraphics[width=0.9\linewidth]{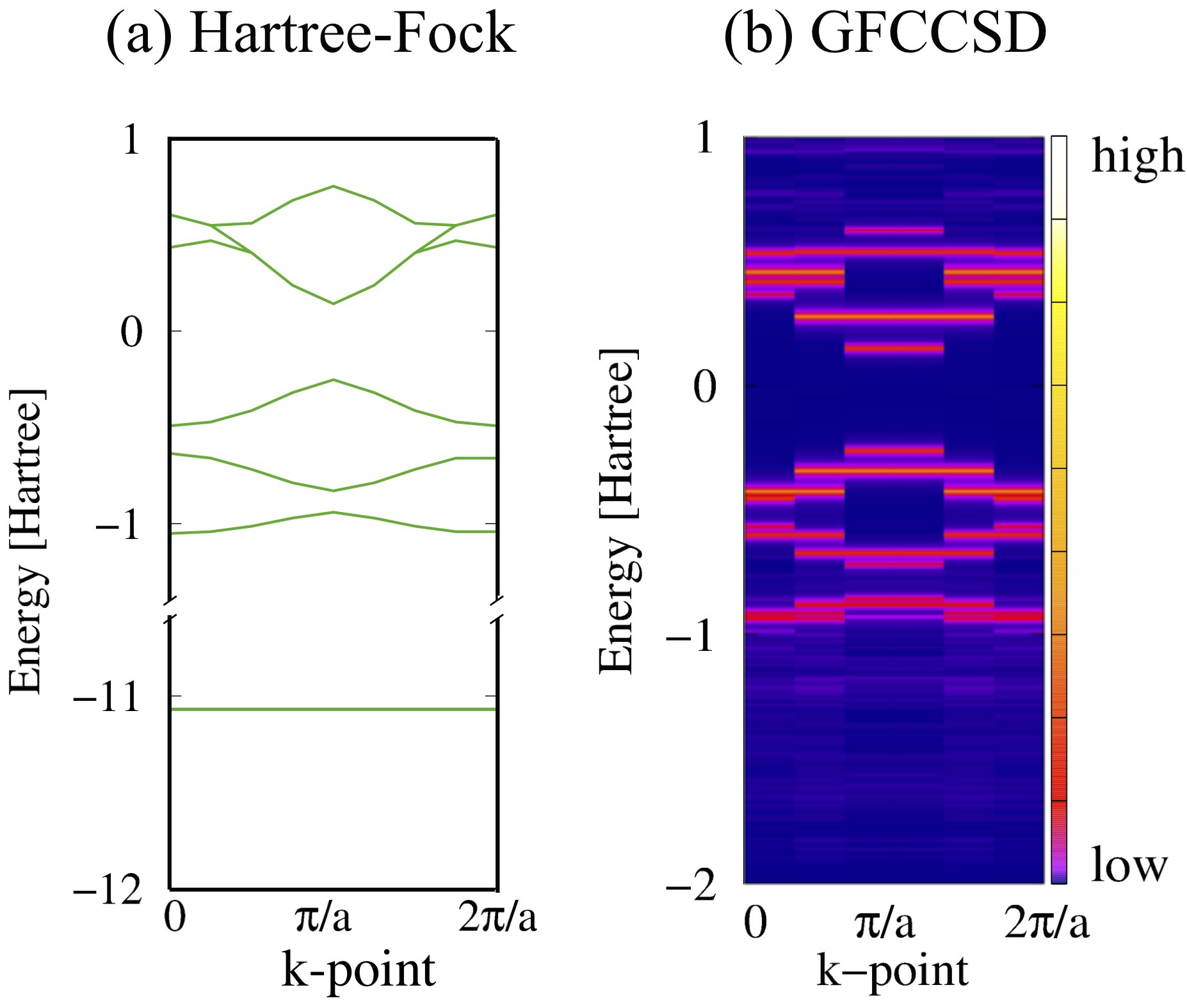}
    \caption{Band structure of the C chain from (a) HF and (b) GFCCSD calculated with $\Nk=6$. $a$ in the horizontal axis represents the lattice constant.}
    \label{img:C-spectrum-all}
\end{figure}

\begin{figure}
    \includegraphics[width=0.5\linewidth]{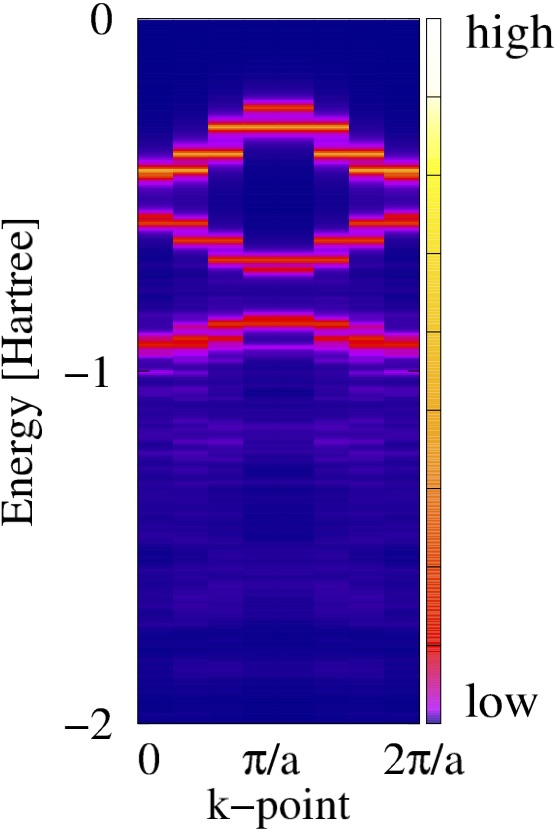}
    \caption{Band structure of the C chain from GFCCSD calculated with $\Nk=8$. $a$ in the horizontal axis represents the lattice constant.}
    \label{img:C-spectrum-k8}
\end{figure}

We first examined the HF band structure of the C-chain. The band structure is shown in Fig.~\ref{img:C-spectrum-all}(a). There are doubly degenerate bands at $-11$ Hartree.
They are found to derive from the $1s$ orbitals of C atoms. The character of the valence-band, which is doubly degenerate, is a hybridized of two carbon $2p$ orbitals perpendicular to the C-C direction. The C chain is a typical covalent material.

Next we explored the possibility of reducing the active space following subsec.~\ref{subsec:active-space}, adopting $\Nk=4$.
The doubly degenerate bands at $-11$ Hartree are expected to make little contribution to the chemical bonding of the system.
Therefore, it is reasonable to exclude these two bands from the active space.
This notion has been found to be valid by confirming that the DOS obtained from all-electron calculation and the one obtained without the deep-level bands coincide with each other near the gap.

Fig.~\ref{img:C-spectrum-all} (b) shows the band structure in GFCCSD with optimized parameters stated above.
The band gap, which is calculated from the peak positions at the Brillouin zone edge $\pi/a$, is 0.50 Hartree, while in HF it is 0.55 Hartree, suggesting the correction of the band structure by the incorporation of the correlation effect.
Also, in this system, we observe satellite peaks below the quasipaticle peak located at $-1$ Hartree. However, one can see a clear difference than those in LiH chain that satellite peaks are much broader than LiH chain.

To take a closer look at the satellite peaks, we show the results from $\Nk=8$ in Fig. \ref{img:C-spectrum-k8}.

\begin{figure}[tb]
    \begin{minipage}[b]{1\linewidth}
        \centering
        \subcaption{Total DOS}
        \includegraphics[width=0.8\linewidth]{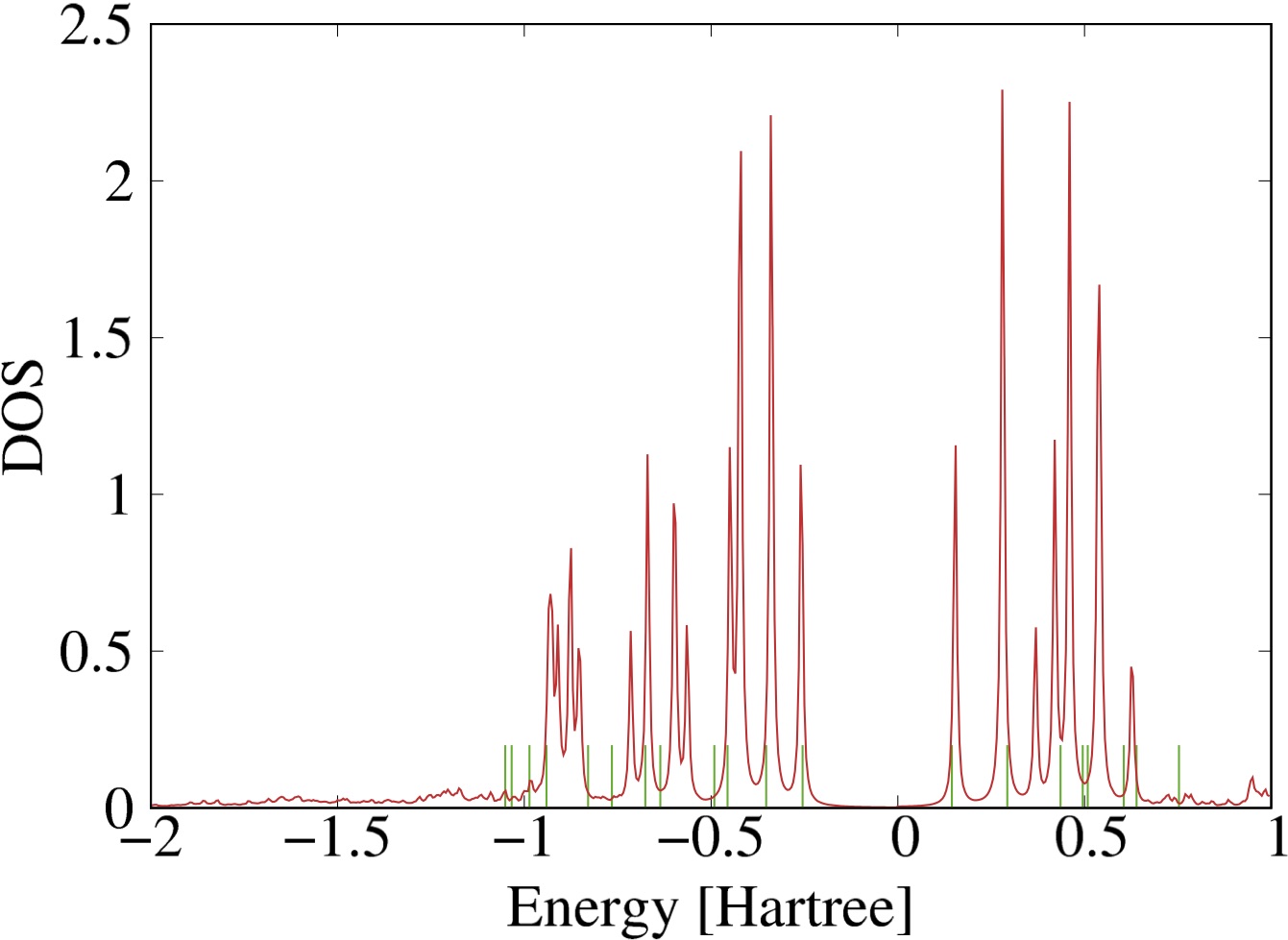}
    \end{minipage} \\
    \begin{minipage}[b]{1\linewidth}
        \centering
        \subcaption{Satellite peaks}
        \includegraphics[width=0.8\linewidth]{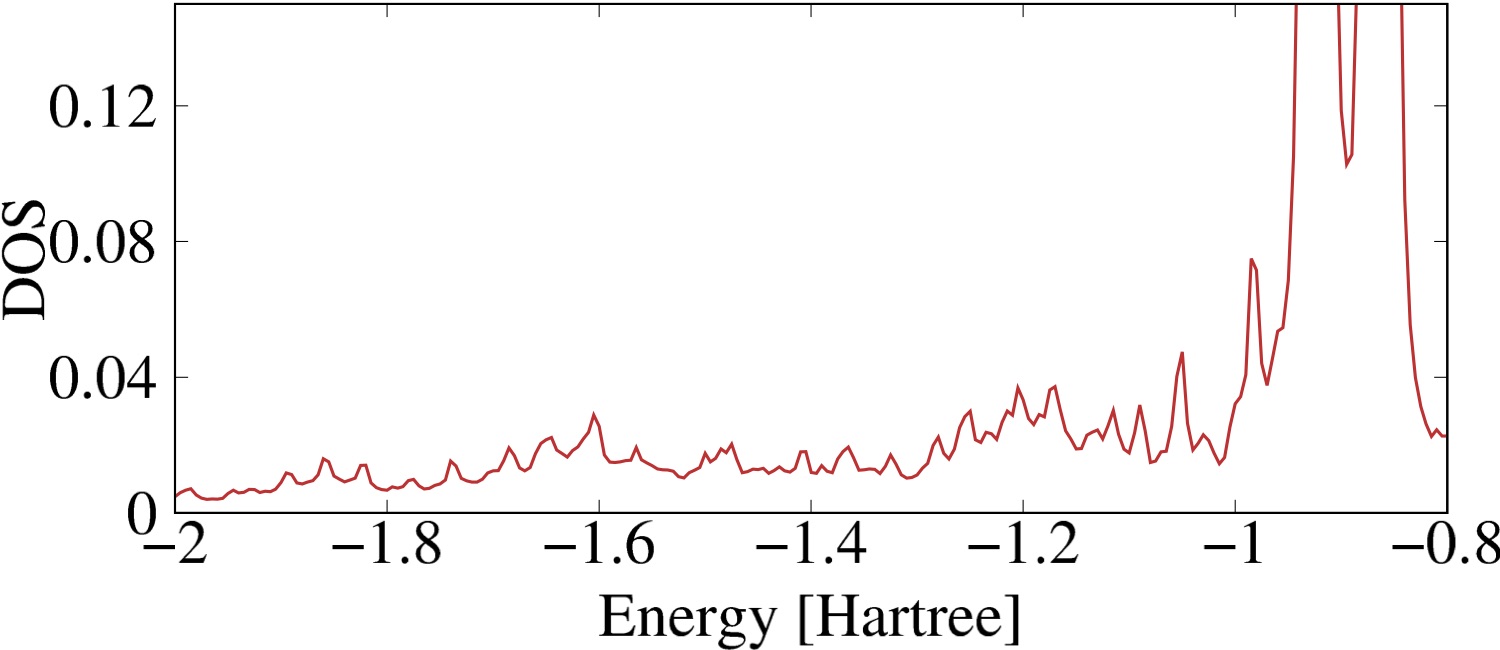}
    \end{minipage}
    \caption{DOS of C chain calculated with $\Nk=6$. The positions of the green sticks in (a) represent the eigenvalues obtained in HF calculations.
    \label{img:C-DOS-k6-cut}}
\end{figure}

The DOS calculated in GFCCSD is shown in Fig.~\ref{img:C-DOS-k6-cut}. Sharp peaks between $-0.96$ and $-0.8$ Hartree and those between $-0.8$ and $-0.5$ come from the second and the third lowest band, respectively, both of which are $sp$-hybridized orbitals that form $\sigma$-bondings with neighboring atoms.
Those between $-0.5$ and $-0.2$, on the other hand, correspond to two degenerate $2p$ orbitals that are orthogonal to the bonding direction and then create $\pi$ bondings.

One distinct feature in the plot is the emergence of broad satellite peaks just below the lowest quasiparticle peak at $-0.8$ Hartree. The integrated value of the satellite peaks, which are located below $-1$ Hartree, is $0.96$.
Considering that this system is spin-degenerate and thus every spacial orbital is occupied by two electrons, this implies that some quasiparticle peaks between -1 and 0 Hartree consist of less than two electrons.
Examining the valence quasiparticle peaks, all of which corresponds to a certain mean-field energy band, we have found that the integration of both the lowest and the second lowest quasiparticle peaks yield 1.5, while those of other peaks are close to 2. This indicates that that the satellite peaks derive the lowest and the second-lowest quasiparticle peaks.

\subsection{One-dimensional Be chain}

\begin{figure}
    \includegraphics[width=0.9\linewidth]{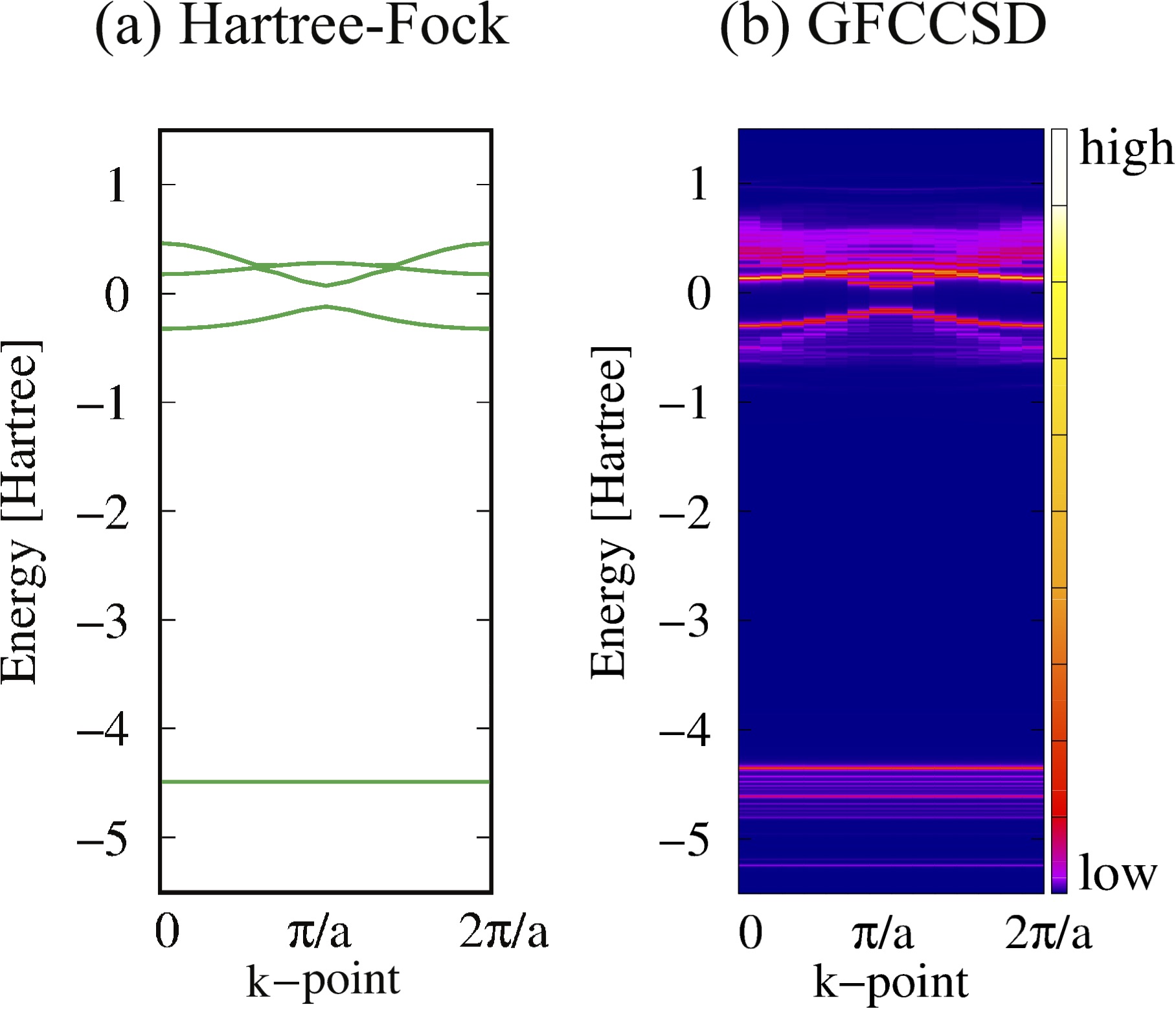}
    \caption{Band structure of the Be chain calculated from (a) HF and (b) GFCCSD with $\Nk=14$. $a$ in the horizontal axis represents the lattice constant.}
    \label{img:Be-spectrum-all}
\end{figure}

\begin{figure}[tb]
    \begin{minipage}[b]{0.45\linewidth}
        \centering
        \subcaption{Deep level}
        \includegraphics[width=1\linewidth]{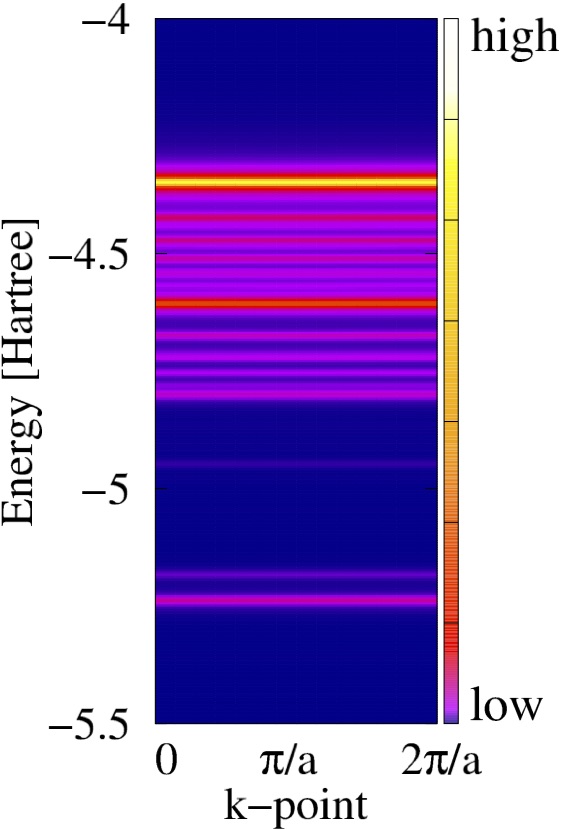}
    \end{minipage}
    \begin{minipage}[b]{0.45\linewidth}
        \centering
        \subcaption{Near the gap}
        \includegraphics[width=1\linewidth]{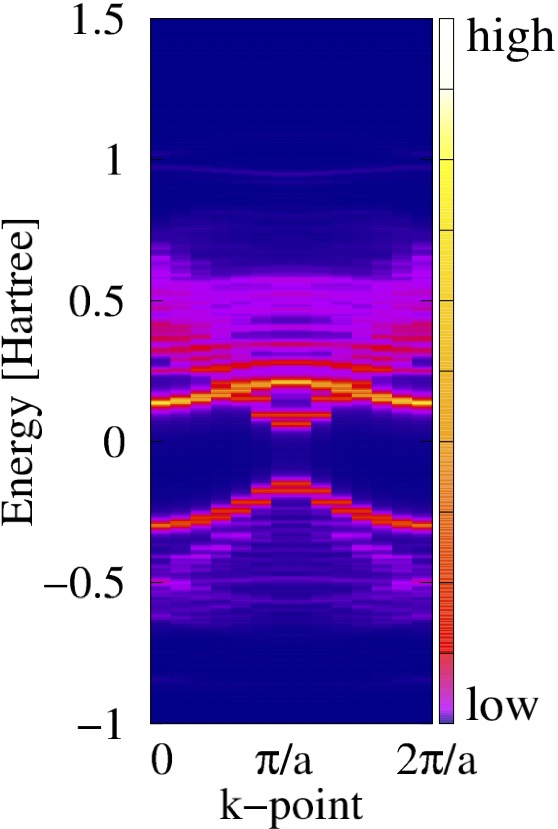}
    \end{minipage}
    \caption{Enlarged illustration of Fig. \ref{img:Be-spectrum-all}: (a) the lowest valence states and (b) those near the gap.
    \label{Be-band-fine}}
\end{figure}

\begin{figure}[tb]
    \includegraphics[width=0.7\linewidth]{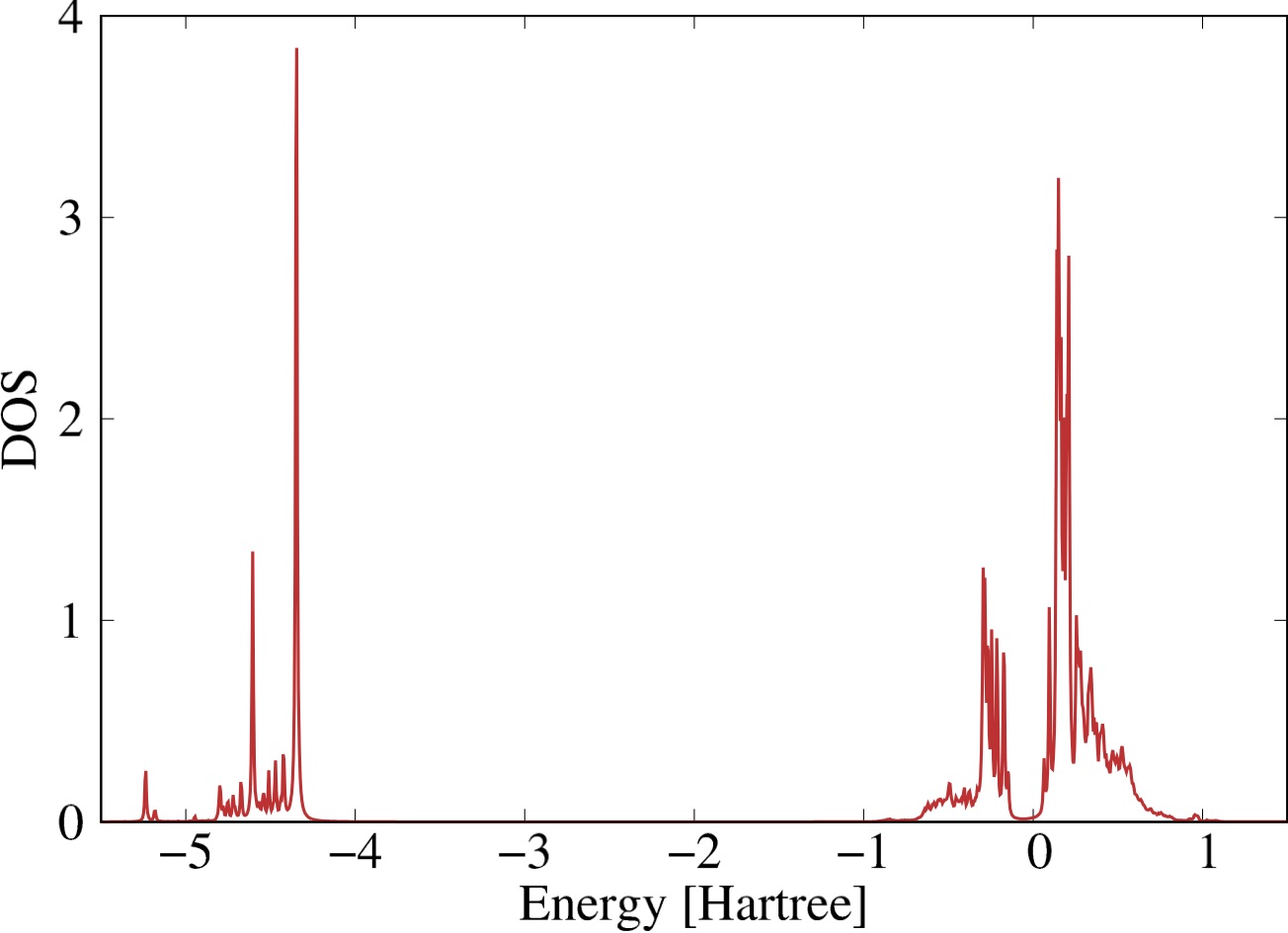}
    \caption{Density of states (DOS) of Be chain calculated from GFCCSD with $\Nk=14$.
    \label{img:Be-DOS-k14-all}}
\end{figure}

As an example of van der Waals materials, we picked up a one-dimension Be chain for our target. Be atom has closed shells up to $2s$ orbital. Therefore, the force condensing the Be atoms is van der Waals interaction.
We used the lattice constant of 3.0 {\AA} which Ref.~\cite{hirata2004} determined. The number of $k$-points is determined to be 14 after checking the accuracy of
the total energy
within the error of 10 meV/cell.
The band structure calculated on HF is shown in Fig.~\ref{img:Be-spectrum-all}(a).
The lowest band located at $-4.5$ Hartree in energy has a character of Be-$1s$ orbital. The second lowest is made up of Be-$2s$ orbital with some dispersion resulting from the interaction between the adjacent Be atoms. In contrast, the lowest conduction band at $\Gamma$ point is doubly degenerate having the Be-$2p$ orbitals perpendicular to the Be-Be direction. The highest band is Be-$2p$ orbital pointing the Be-Be direction.

We applied the GFCCSD method to this system, whose result is shown in Fig.~\ref{img:Be-spectrum-all}(b).
The overall features of quasiparticle peaks are understood by comparing with the HF results.
One of the most interesting points is the appearance of two different kinds of satellite peaks. One can see discrete and almost flat satellite peaks (see also Fig.~\ref{Be-band-fine}(a), in which the satellite peaks around $1s$ are shown in the enlarged picture).
We have checked the dependency of the number of satellite peaks on $\Nk$ by changing $\Nk$ from 10 to 14. We did not find, however, any differences in the number of satellite peaks.
Therefore we conclude that the number of the satellite peaks is completely independent of the number of $k$-points.
The similar satellite peaks are also seen in unoccupied side above $0$ Hartree. The doubly degenerate conduction band has many duplicates above the quasiparticle bands.
The other type of satellite peaks is observed below the highest valence band in the energy region between $0$ and $-0.7$ Hartree. For the satellite peak, we cannot see duplicate bands different from the other ones.
By increasing $\Nk$, we can see a band structure of satellite peaks below the highest valence band with different dispersions of the valence band.
The calculated DOS is also shown in Fig.~\ref{img:Be-DOS-k14-all}.
We can also see the differences in the two kinds of satellite peaks in the figure: one that appears like a spiky structure, and the one that is broaden or bump peak structure.

\section{Conclusion} \label{conclusion}
We have calculated the band structures through GFCCSD method for various kinds of systems from ionic to covalent and van der Waals systems for the first time: one-dimensional LiH chain, one-dimensional C chain, and one-dimensional Be chain.
We have found that the band gap becomes narrower than in HF due to the correlation effect.
We have also shown that the band structures obtained from GFCCSD, which includes both quasiparticle and satellite peaks.
Also, taking one-dimensional LiH as an example, we have discussed the validity of restricting the active space to suppress the computational cost of GFCCSD while keeping the accuracy, and found that the calculated results without bands that do not contribute to the chemical bondings were in good agreement with full-band calculations.
By GFCCSD method, we can calculate the total energy and band structures within the framework of CCSD with a great accuracy.

\begin{acknowledgments}
This research was supported by MEXT as Exploratory Challenge on Post-K computer'' (Frontiers of Basic Science: Challenging the Limits). This research used computational resources of the K computer provided by the RIKEN Advanced Institute for Computational Science through the HPCI System Research project (Project ID: hp170261). Y.M. acknowledges the support from JSPS Grant-in-Aid for Young Scientists (B) (Grant No. 16K18075).
\end{acknowledgments}

\bibliographystyle{apsrev4-1}
\bibliography{paper}


\end{document}